\documentclass{article}

\usepackage{blindtext}
\usepackage{multicol} 
\usepackage{caption}
\usepackage{subcaption}
\usepackage{graphicx} 
\graphicspath{ {./images/} }
\usepackage[a4paper, total={7in, 9in}]{geometry}
\usepackage{url}
\usepackage[utf8]{inputenc}
\usepackage[english]{babel}
\usepackage{amsmath, amsfonts, amssymb}
\usepackage{csquotes}
\usepackage{float} 
\usepackage{authblk}
\usepackage[
  natbib=true,
  style=numeric,
  sorting=none
]{biblatex}
\usepackage{hyperref}

\usepackage{soul, color}

\newcommand{\UNIFI}{Dipartimento di Fisica, Universit\`a degli Studi di Firenze, via Sansone 1, Sesto Fiorentino, Italy}
\newcommand{\CNRINO}{CNR-INO, Largo Enrico Fermi 6, Firenze, Italy}
\newcommand{\QTI}{QTI srl, Largo Enrico Fermi 6, Firenze, Italy}
\newcommand{\GGI}{The Galileo Galilei Institute for Theoretical Physics (GGI)}

\title{Time-bin encoding quantum key distribution in free-space horizontal links during nighttime and daytime}
\author[1,2]{Sebastiano Cocchi}
\author[1,2]{Domenico Ribezzo}
\author[1,2]{Giulia Guarda}
\author[4]{Pietro Centorrino}
\author[3]{Tommaso Occhipinti}
\author[2]{Alessandro Zavatta}
\author[1,2,3 *]{Davide Bacco}

\affil[1]{\UNIFI}
\affil[2]{\CNRINO}
\affil[3]{\QTI}
\affil[4]{\GGI}
\affil[*]{davide.bacco@unifi.it}
\date{}                    
\begin{document}

\maketitle

\begin{abstract}

Free-space quantum key distribution (QKD) represents a groundbreaking advancement in secure communication, enabling secure key exchange over vast distances and offering robust encryption for the future quantum internet.
However, the compatibility between fiber and free-space infrastructures continues to pose challenges for QKD protocols.
Indeed, free-space and fiber-based networks commonly use different wavelengths and qubits encoding schemes.
On the one hand, free-space QKD typically exploits visible light for its beneficial beam divergence compared to longer wavelengths, and polarization encoding for its robustness against turbulence.
On the other hand, fiber-based QKD employs infrared light, particularly the C-band, because it shows the minimum losses with silica fibers, and time-bin encoding, due to polarization instability in optical fibers.
In our study, we demonstrate the viability of a time-bin encoded QKD protocol operating in the C-band through horizontal turbulent free-space channels.
We test the setup into a 50\ m and a 500 m long links, achieving an average secure key rate of, respectively, 793 kbps and 40 kbps over several hours of measurements. 
The results encourage further exploration of the interoperability between free-space and fiber-based infrastructures, opening new possibilities for connecting terminal users with satellites in hybrid infrastructures.

\end{abstract}

\begin{multicols}{2}

\begin{figure*}[t!] 
  \begin{center}
    \includegraphics[width=0.99\textwidth]{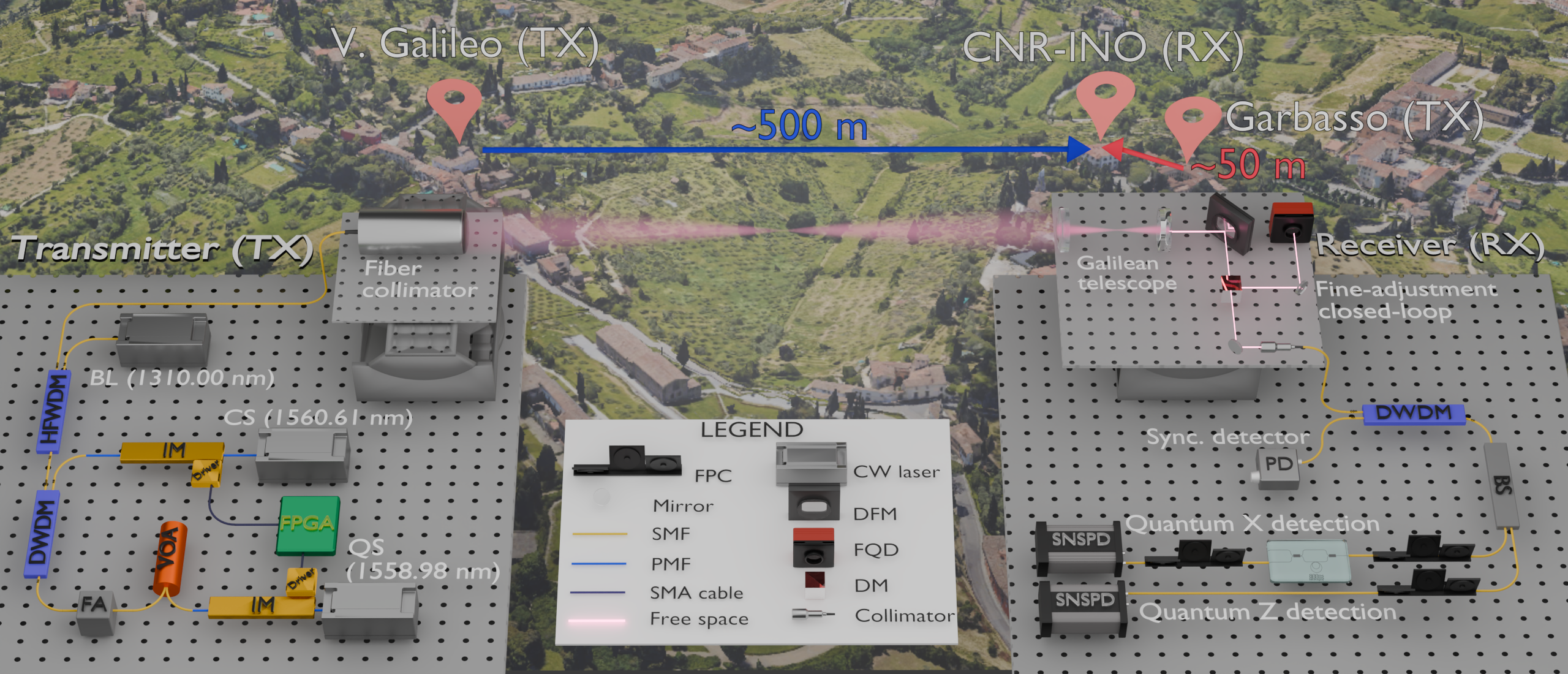}
  \end{center}
    \caption{\textbf{Experimental setup scheme.}
    The map shows the two links (photo credits: Google Maps).
    The map's point of view is facing the southwest direction.
    The red arrow displays the 50 m long link between the Garbasso and the CNR-INO.
    The blue arrow represents the 500 m long link between Villa Galileo and the CNR-INO.
    The bottom of the figure shows the setup design.
    FPGA: field programmable gate array;
    IM: intensity modulator;
    VOA: variable optical attenuator;
    FA: fixed attenuation;
    DWDM: dense wavelength division multiplexer;
    HFWDM: high-frequency wavelength division multiplexer;
    SMF: single-mode fiber;
    PMF: polarization maintaining fiber;
    DFM: deformable mirror;
    FQD: four-quadrant detector;
    DM: dichroic mirror;
    CW: continuous wave;
    PD: photodiode;
    FPC: fiber polarization controller.}\label{fig:experimental-setup}
\end{figure*}

\section{Introduction}
\label{sec:introduction}

Interest in free-space (FS) optical quantum communications has been growing annually, intending to establish a global quantum internet network \cite{Simon2017, Conti2024SatelliteTerrestrialQN}.
For example, the number of satellites developed by China, Europe, and other players for quantum key distribution (QKD), a provable secure method to share private keys \cite{Bennett_84, Ekert1992, Scarani_2009, pirandola2020advances}, is rapidly growing \cite{SAGA_1, Yin2020, Ntanos2022LargeS}.
However, despite continuous efforts, achieving interoperability between fiber infrastructures and FS links remains a significant challenge for QKD setups.

FS QKD and fiber-based QKD cannot directly exchange qubits primarily because they rely on different wavelengths and encoding schemes.
FS QKD typically utilizes visible light, taking advantage of its beneficial beam divergence compared to the infrared (IR) light\cite{Saleh:1084451}.
It generally employs polarization encoding to enhance robustness against atmospheric turbulence \cite{Cai:27, Liao2017, hughes2002practical, cao2020long, Yin2020, Berra2023, Liao2017}.
In contrast, fiber-based QKD operates in the IR spectrum, particularly the C-band, which minimizes losses in silica optical fibers.
Moreover, fiber-based QKD generally utilizes time-bin and phase encoding for its robustness against polarization drift over the fiber channel \cite{ribezzodeploying, Dynes2019, refId0, Neumann2022, PhysRevA.98.052336}.
At present, sharing a quantum signal from an optical fiber to a free-space link typically requires a third-party device, which adds an unnecessary insecurity layer to the entire communication.
A possible solution to achieve full interoperability between fiber-infrastructure users and satellites relies on the same QKD implementation, \textit{i.e.} qubit encoding and chosen wavelength.

In this work, we employ time-bin encoding in the C-band over multiple FS QKD links.
C-band is a favorable choice over visible and near-infrared light since it reduces the turbulence-induced blurring effect \cite{Fried:66, Ciddor:96}.
Furthermore, the sun's spectral irradiance at 1550 nm is approximately five times lower than at 800 nm and the Rayleigh scattering is only $7\,\%$ compared to the one at 800 nm \cite{Liao2017LongdistanceFQ}.
Instead, time-bin encoding is generally considered disadvantageous in a turbulent FS channel due to phase instabilities \cite{PhysRevA.97.043847}.
However, a few studies have explored its implementation in horizontal QKD applications with visible light over FS channels with phase compensation at the receiver side, showing promising results \cite{Jennewein:23, Jin:19}.
As evidence of this, the European Union's SAGA project will deploy a satellite equipped with a phase-encoded QKD transmitter in the C-band for demonstrative applications \cite{SAGA_1}.

In our experiment, we showcase a typical fiber-based approach in an FS QKD apparatus over two different horizontal FS channels of $50\,$m and $500\,$m.
We reach an average secure key rate (SKR) of, respectively, $793\,$kbps, and $40\,$kbps over several hours during both daytime and nighttime.
This is enabled by coupling the quantum signal into single-mode fiber (SMF), which is robust against spurious light coming from the environment, for example, the sunlight \cite{Takenaka:12}.
We also exploit a photonic integrated circuit (PIC) imbalanced Mach-Zender interferometer (IMZI) as part of the decoding setup, allowing stable visibility and minimal losses \cite{PhysRevA.110.042605}.
Ultimately, our approach offers a dependable solution that seamlessly connects urban fiber users with FS optical channels, paving the way for integrating quantum communications with classical telecommunications infrastructures, including satellite intercontinental links.

\section{Experiment}
\label{sec:experiment}

We implement a three-state efficient one-decoy BB84 QKD protocol \cite{PhysRevA.98.052336,boaron2018simple,lo2005decoy} in two metropolitan FS optical links in Florence (Italy).
Details about the implementation and the protocol are explained in the Supplementary materials.
The first link, consisting of approximately \(50\,\)m, connects the Garbasso building, part of the physics department at the University of Florence, with the headquarters of the National Institute of Optics of the National Research Council (CNR-INO).
The second quantum link, extending to $500\,$m, connects Villa il Gioiello -- Galileo Galilei's last residence, now a university museum -- with the CNR-INO.
All the necessary instruments for quantum detection are located at the CNR-INO.

The transmitter (Alice) generates three optical signals: the quantum signal (QS), a clock signal (CS) for synchronization purposes, and a beacon laser (BL), as depicted in Figure \ref{fig:experimental-setup}. 
The QS is a sequence of time-bin-encoded quantum states generated with a rate of 595 MHz and with a delay between \textit{early} and \textit{late} states of \(\tau=800\,\)ps.
To produce these states, an intensity modulator (IM) carves the light coming from a continuous wave (CW) laser with a wavelength of \(1558.98\,\)nm. 
The electrical signal that powers the IM is generated by a field-programmable gate array (FPGA).
A variable optical attenuation (VOA) stage, along with some fixed attenuation (FA), enables the achievement of single-photon intensity.

Alice and Bob need to synchronize their clock to assign the correct timestamp to each quantum state.
In the experiment, we use a clock-over-the-air signal, the CS, to accomplish this task.
After the signals' generation, a dense wavelength division multiplexing device (DWDM) combines the \(1560.61\,\)nm CS with the QS.

The BL is a \(1310.10\,\)nm continuous wave laser that we use as a reference to compensate for the beam wandering effect at the receiver side.
Then, a high-frequency WDM (HFWDM) combines the BL with the CS and the QS.
A custom motorized platform hosts the transmitter optical system which can be moved along the degrees of freedom of altitude and azimuth (ALTAZ).
The ALTAZ mounts are practical for the coarse alignment since we can move them remotely with adequate precision.
Finally, a variable-focus fiber collimator delivers the three signals to the receiver through the FS channel.

The receiver is divided into an FS and a fiber-based section. 
We place the free-space section on the upper floor of the CNR-INO upon a second ALTAZ mount.
The incoming signals are collected by a Galilean telescope with an outer lens diameter of \(35\,\)mm and a magnification factor of $10$. 
Right after, the tip and tilt deformable mirror (DFM) reflects the beam toward a dichroic mirror (DM).
The DM separates the BL from the other signals and a four-quadrant detector (FQD) measures the beacon position error.
A proportional-integrative-derivative (PID) controller adjusts the DFM position, closing the fine-adjustment loop. 
With this correction, we optimize the coupling of the QS and the CS into the SMF.

The QS and CS are transmitted to the quantum detection apparatus which is placed in the laboratory.
Here, the CS and the QS are demultiplexed by a DWDM.
On the CS path, we retrieve the analog clock using a photodiode (PD) and we use it to synchronize the time-tagger unit.

In the quantum receiver, a beam splitter (BS) with a 50:50 ratio acts as Bob's random basis choice.
A superconducting nano-wire single-photon detector (SNSPD) measures the arrival times of photons directly from the first output port, which represents the Z-basis detection.
The second output port is routed towards the PIC-based IMZI for detection in the X-basis.
A delay line of \(\tau=800\,\)ps in one arm ensures the correct phase measurement.
Another SNSPD measures the photon clicks coming from one IMZI output.
Then, a time-tagger unit converts each SNSPD signal into digital data, which we process using our software.
The software calculates the Quantum Bit Error Rate (QBER) for the Z-basis and the visibility, thus the QBER, for the X-basis, as shown in Supplementary materials.
Finally, we calculate the secure key rate (SKR) in the finite-key regime for a block size of $n_Z=10^7$ qubit exchanged with the measured $QBER_Z$ and $QBER_X$.

\section{Results}
\label{sec:results}

\begin{figure*}
\begin{subfigure}[b]{0.5\textwidth}
    \centering
    \includegraphics[width=\textwidth]{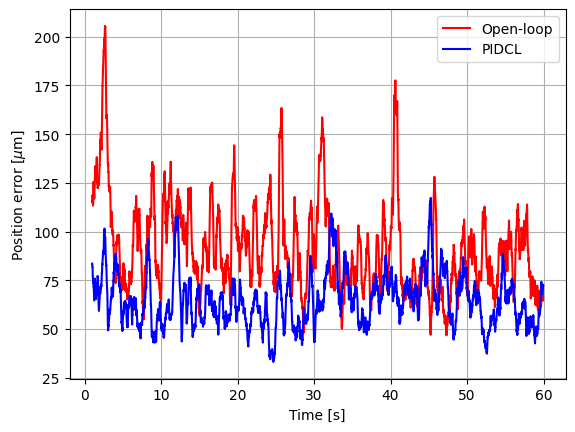}
    \caption{}
  \label{fig:results-pid}
\end{subfigure}%
\begin{subfigure}[b]{0.5\textwidth}
    \centering
    \includegraphics[width=\textwidth]{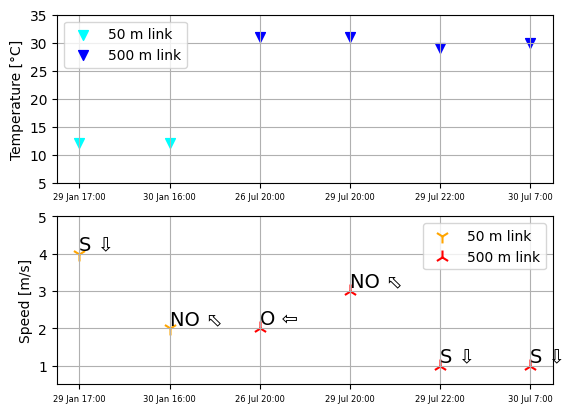}
    \caption{}
  \label{fig:results-weather}
\end{subfigure}
\begin{subfigure}[b]{0.5\textwidth}
    \centering
    \includegraphics[width=1.1\textwidth]{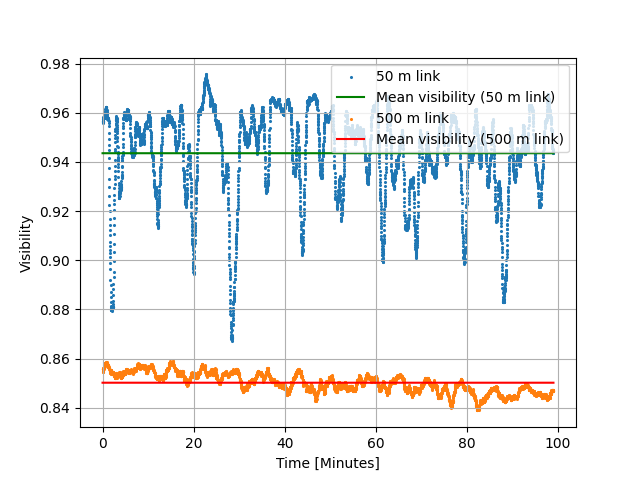}
    \caption{}
  \label{fig:results-vis}
\end{subfigure}%
\begin{subfigure}[b]{0.5\textwidth}
    \centering
    \includegraphics[width=\textwidth]{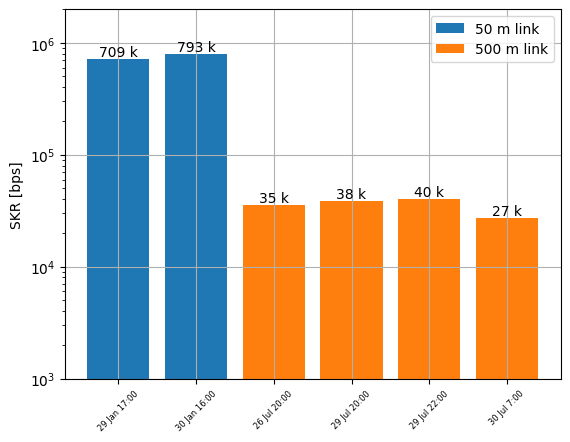}
    \caption{}
  \label{fig:results-skr}
\end{subfigure}
\caption{\textbf{Experimental results}.
    a) Comparison between the open-loop and closed-loop configurations of the fine-adjustment loop. The position errors are taken from the FQD in the open-loop (red line), that is without fine-adjustment, and with the closed-loop configuration (blue line). Data come from the $500\,$m link. 
    b) Weather conditions during the QKD communications. The first triangular cyan measurements refer to the $50\,$m link, and the other points to the $500\,$m communication. 
    c) Comparison of the visibility stability for two hours between the $50\,$m long link (blue dots) and the $500\,$m long link (orange dots). Each point is 60 seconds.  
  d) Achieved SKR. The first two bars refer to the $50\,$m link, and the other bars to the $500\,$m communication.}
    \label{fig:results}
\end{figure*}

As a first result, we report the estimate of the mean FS channel attenuation during the QKD experiments.
The temperature difference between the rooms and the external environment causes convective motions, leading to an increase in the turbulence effect.
In order to reduce this effect, and to comply with the regulation of our Department, we chose to close the windows of the two buildings.
This action drastically reduces the turbulence effect, at a cost of an additional $\sim$3 dB of losses.
For the $50\,$m experiment we measure an average channel loss of $7\,$dB, basically due to the windows glasses and the SMF coupling losses. 
The $500\,$m communication shows an average channel loss of 16--17$\,$dB. 
In this case, the losses are primarily coupling losses due to beam divergence.

We also characterize the turbulence strength of the two channels by estimating the scintillation index, as reported in Supplementary materials.
We compute a scintillation index of $\sigma_I^2 = 3.1\times10^{-5}$ for the $50\,$m link and $\sigma_I^2 = 2.12\times10^{-4}$ for the $500\,$m link.
Since $\sigma_I^2<1$ for both links, we conclude that we operate in a weak turbulence regime.

Then, we compare the performances between the open-loop configuration and the PID controller closed-loop (PIDCL), as shown in Fig. \textbf{\ref{fig:results-pid}}. 
The open-loop configuration has a mean position error of \(92\,\mu\)m with a standard deviation of \(53\,\mu\)m.
The PIDCL scored a mean position error of \(65\,\mu\)m with a standard deviation of \(36\,\mu\)m. 
The PIDCL configuration exhibits not only a lower coupling misalignment but also a lower degree of fluctuation compared to the open-loop configuration, as reflected by its lower standard deviation.

Finally, we perform the acquisition of the quantum bits over several hours of measurements for both links in different weather conditions.
In Fig. \textbf{\ref{fig:results-weather}}, we report the temperature and the wind speed for the different trials of the quantum acquisition.
In Fig. \textbf{\ref{fig:results-vis}}, we display the visibility stability over approximately 2 hours of measurements.
The mean values for the visibility are $94\%$ and $85\%$ for the 50 m link and the 500 m link.
The data for the 50 m link starts from 5 pm CET and is characterized by a consistent fluctuation over time.
This fluctuation is attributed to the optical transmitter's location, which is installed in a working environment where people frequently walk around.
The data for the 500 m link was collected starting at 8 pm CET, with a transmitter placed in a temporarily closed location with no human activity.

Additionally, we report in Fig. \textbf{\ref{fig:results-skr}} the SKR achieved in the 50 m link (blue rectangle) and 500 m link (orange rectangles) links in various daytime and nighttime scenarios.
We selected a block size of $10^7$ because it can be collected within a few seconds across both links.
The mean SKR achieved in the 50 m link ranges from 709 kbps to 793 kbps, while the mean SKRs in the 500 m link are 35 kbps, 38 kbps, and 40 kbps.

\section{Discussion}
\label{sec:discussion}

In this section, we compare our results with the latest works in the existing literature.
It is worth noting that direct comparisons of achieved distances and rates are challenging because the adopted FS optics strongly affect channel attenuation.
Furthermore, the interferometer's visibility degrades with increasing FS distances, making estimations of the SKR in lossier channels particularly tricky.
In \cite{Jin:19}, the reported average SKR is 154 bps using a time-bin encoded weak coherent source at 150 MHz in a 1.2 km link with 38 dB attenuation.
In contrast, our setup achieved a better alignment, with approximately 20 dB of the overall link budget measured in a 500-meter link.
In a similar channel with 38 dB loss, we estimate that our setup could reach up to 400 bps.
In another study, \cite{Bulla2022NonlocalTI} achieved an SKR of 100--200 bps in a 25 dB attenuation channel over 10 km using a polarization-encoded entangled-photons source.
Our setup, on the other hand, demonstrated an average SKR of 40 kbps over a 16 dB channel.
With optimized optics, we expect to achieve an SKR of 4 kbps in a 25 dB channel.
Our improved results can be attributed to our faster qubit generation rate of 595 MHz and our high-stability IMZI adopted for visibility readout.

As reported in the previous section, the losses for the $500\,$m free-space link are significantly higher ($16\,$dB) compared to the $50\,$m link ($7\,$dB).
The difference of over 10 dB between the two channels, not attributable to the atmospheric loss of a 500 m channel, is due to undersized optics.
Indeed, according to Gaussian beam propagation principles \cite{Saleh:1084451}, the radius of a $1550\,$nm Gaussian beam expands from an initial value of $\omega_0=7\,$mm to $\omega_{1}=35\,$mm after traveling a distance of $500\,$m.
Given $D=35\,$mm, optical aperture of the Galilean telescope on the receiver side, the telescope truncates a portion of the incoming beam, as it has a diameter of $d_1=2 \omega_1=70\,$mm.

In addition, the optical alignment between the two ALTAZ mounts deteriorates over time.
The thermal characteristics of the materials used in the mounts can expand or contract with temperature fluctuations up to 15° C of excursion between day and night during the summer \cite{lamma}.
We plan to introduce a coarse alignment system to preserve the mounts' alignment during extended communication periods.
Similarly, we devise the adoption of larger optical transmitter and receiver stations, reducing the beam divergence and facilitating longer distances in future links.

One final improvement could be employing extensive Artificial Intelligence (AI) in the closed-loop, as we proposed in \cite{cocchi:AI}.
AI can optimize the PID parameters to fit the current turbulence strength and could help in improving the overall stability of the system.

\section{Conclusions}
\label{sec:conclusions}

We demonstrated the practicability of QKD implementation with time-bin encoding in the C-band over two distinct horizontal FS links.
In the first link, ranging up to $50\,$m, we reach a mean SKR of $793\,$kbps;
in the second link, we achieve a mean SKR of $40\,$kbps with good stability over two hours.

With its simplicity, elevated qubit generation rate, and easy integration with fiber infrastructures, our setup demonstrates the possibility of using time-bin encoding in the C-band for multi-domain quantum networks.
This development is a significant step towards achieving high interoperability between different network infrastructures, which will play a crucial role in building the future global quantum network.

\section{Funding}
This work was funded by the European Union (ERC, QOMUNE, 101077917, by the Project EQUO (European QUantum ecOsystems) which is funded by the European Commission in the Digital Europe Programme under the grant agreement No 101091561, the Project SERICS (PE00000014) under the MUR National Recovery and Resilience Plan funded by the European Union - NextGenerationEU, the Project QuONTENT under the Progetti di Ricerca, CNR program funded by the Consiglio Nazionale delle Ricerche (CNR) and by the European Union - PON Ricerca e Innovazione 2014-2020 FESR - Project ARS01/00734 QUANCOM, the Project QUID (Quantum Italy Deployment) funded by the European Commission in the Digital Europe Programme under the grant agreement No 101091408.

\printbibliography

\newpage
\section{Supplementary materials}
        
\subsection{The QKD protocol}
\label{sec:supplementary-protocol}
In the prepare and measurement QKD scheme, the transmitter (Alice) encodes the key bits into qubits using, at least, two mutually unbiased bases.
The receiver (Bob) randomly chooses which basis to measure each qubit. 
In the case of time-bin and phase encoding, the two bases (Z and X) are encoded in the time difference between two pulses.
For the Z-basis, if a photon arrives in the first time-bin, it is referred to as \textit{early} (E), while arrival in the second adjacent time-bin is referred to as \textit{late} (L). 
In Dirac notation, one can represent these states as $|E\rangle$ and $|L\rangle$.
The X-basis contains the two states whose wavefunction is the overlapping of E and L with two possible phases, namely \(0\) and \(\pi\) 
\begin{align}
    |+\rangle &= \frac{|E\rangle+e^{i\varphi}|L\rangle}{\sqrt{2}}\Bigg\vert_{\varphi=0} = \frac{|E\rangle+|L\rangle}{\sqrt{2}},\\
    |-\rangle &= \frac{|E\rangle+e^{i\varphi}|L\rangle}{\sqrt{2}}\Bigg\vert_{\varphi=\pi}= \frac{|E\rangle-|L\rangle}{\sqrt{2}}.
\end{align}

For the three-states efficient BB84 with one-decoy implementation \cite{boaron2018simple, PhysRevA.98.052336}, the key length that Bob can extract in the finite key regime is \cite{Boaron2018SecureQK}
\begin{align}
    l \leq & s^l_{Z,0} + s^l_{Z,1}(1-H_2(\phi_Z)) - \lambda_{EC} \nonumber \\ 
    &- 6\log_2\left(\frac{19}{\epsilon_{sec}}\right) - \log_2\left(\frac{2}{\epsilon_{corr}}\right)
    \label{eqn:key-length}
\end{align}
where $s^l_{Z,0}$ and $s^l_{Z,1}$ are the vacuum and the single-photon events lower bounds, $H_2$ is the binary entropy, $\phi^{\mu}_Z$ is the phase error rate upper bound, proportional to $QBER_X$, $\lambda_{EC}=s^l_{Z,1}\cdot f_{eff} \cdot H_2(QBER_Z)$ is the number of bits revealed during the error correction process, with $f_{eff}>1$ being the error reconciliation efficiency, and $\epsilon_{sec}$ and $\epsilon_{corr}$ are the secrecy and correctness parameters.

\subsection{Turbulence}
\label{sec:supplementary-turbulence}

Particulates in suspension inside the atmosphere, water vapor, CO\(^2\), fog, \textit{etc.} contributes to the channel attenuation, which cannot be tackled.
Additionally, the different concentrations of the various gases along with temperature fluctuations and wind, contribute to local variations of the air refractive index \cite{Edlen_1966, Birch_1993, kolmogorov1991dissipation}. 
In the atmosphere, the movement of these regions of air, the so-called eddies or vortices, originates the turbulence effects. 
For a small optical system on the order of a few centimeters, like the one we use in the field trial, the more relevant turbulence effects are the beam wandering and the angle of arrival fluctuations \cite{Ciddor:96}. 

The fluctuation of the optical power around a mean value caused by the beam traveling through a turbulent channel is called scintillation \cite{Fried:66, Ciddor:96}.
The structure parameter $C_n^2$ along with the scintillation index $\sigma_I^2$ are a measure of the turbulence strength. 
The scintillation index can be estimated by 
\begin{equation}
    \sigma_I^2 = \frac{\langle I^2\rangle - \langle I\rangle^2}{\langle I\rangle^2} = \frac{\langle I^2\rangle}{\langle I\rangle^2}-1,
  \label{eq:scintillation-index}
\end{equation}
where $I$ is the beam intensity.
The \eqref{eq:scintillation-index} defines two turbulent regimes: (a) the weak irradiance fluctuations when $\sigma_I^2<1$ and (b) the moderate-to-strong irradiance fluctuations when $\sigma_I^2\geq 1$.
One can measure the $C_n^2$ in a horizontal path following log-intensity $\sigma_{ln, I}^2$ \cite{Dix-Matthews:23} 
\begin{equation}
  \sigma_{ln,I}^2 = \ln\left(1 + \frac{\sigma^2}{\mu^2}\right)
  \label{eq:cn2-sigma}
\end{equation}
where $\sigma$ and $\mu$ are, respectively, the standard deviation and the mean of beam optical power. 
For a horizontal link, the log intensity reads 
\begin{equation}
  \sigma_{ln,I}^2 = 0.496 C_n^2 k^{7/6} L^{11/6}
  \label{eq:cn2-log-intensity}
\end{equation}
where $k=2\pi/\lambda$ is the wave number and $\lambda$ the wavelength.
Combining \eqref{eq:cn2-sigma} and \eqref{eq:cn2-log-intensity} we arrive at
\begin{equation}
  C_n^2 =\frac{ \ln\left(1 + \frac{\sigma^2}{\mu^2}\right)}{0.496 k^{7/6} L^{11/6}}
  \label{eq:cn2-method-2}
\end{equation}
The Fried parameter $r_0$ is another relevant quantity characterizing the turbulence affecting the channel. 
It tells that the limiting resolution of the turbulent channel, characterized by a structure constant $C_n^2$, using a $\lambda$ wavelength optical wave with curvature $R$ with infinite diameter pupil is the same as if the receiver collects the same signal in a vacuum distance of same length $L$ with a diffraction-limited lens of diameter $r_0$.
For a horizontal link, the Fried parameter is   
\begin{equation}
  r_0^{hor}  = \left( 1.46 k^2 C_n^2 L \right)^{-3/5}
  \label{eq:fried-parameter}
\end{equation}

We give an estimate of the turbulence strength occurring in the two channels using the scintillation index $\sigma_I^2$ defined in equation \eqref{eq:scintillation-index}, the structure parameter $C_n^2$ defined in \eqref{eq:cn2-method-2}, and the Fried parameter $r_0$ in \eqref{eq:fried-parameter}.
We used the beacon laser as a light source and the FQD to measure the beam's intensity $I$ fluctuations. 
Using the equation \eqref{eq:scintillation-index}, we compute a scintillation index of $\sigma_I^2 = 3.1\times10^{-5}$ for the 50 m long link and $\sigma_I^2 = 2.12\times10^{-4}$.
Since $\sigma_I^2<1$ for both links, we conclude that the turbulence is weak. 
The $C_n^2$ and the Fried parameter confirm this conclusion.
For the $50\,$m experiment we found the following results $C_n^2 = 2.3\times10^{-18}\,$m$^{-2/3}$ and $r_0^{hor}=7.08\,$m.  
For the 500 m long experiment we measure $C_n^2 = 7.71\times 10^{-17}\,$m$^{-2/3}$ and $r_0^{hor}=8.5\times 10^{-1}\,$m.  
In both links $r_0$ is greater than the optical receiver's aperture $D=3.5\times 10^{-2}\,$m, thus confirming the weak turbulence regime.

\subsection{Methods}
\label{sec:supplementary-methods}

\begin{figure*}
    \centering
    \includegraphics[width=0.99\linewidth]{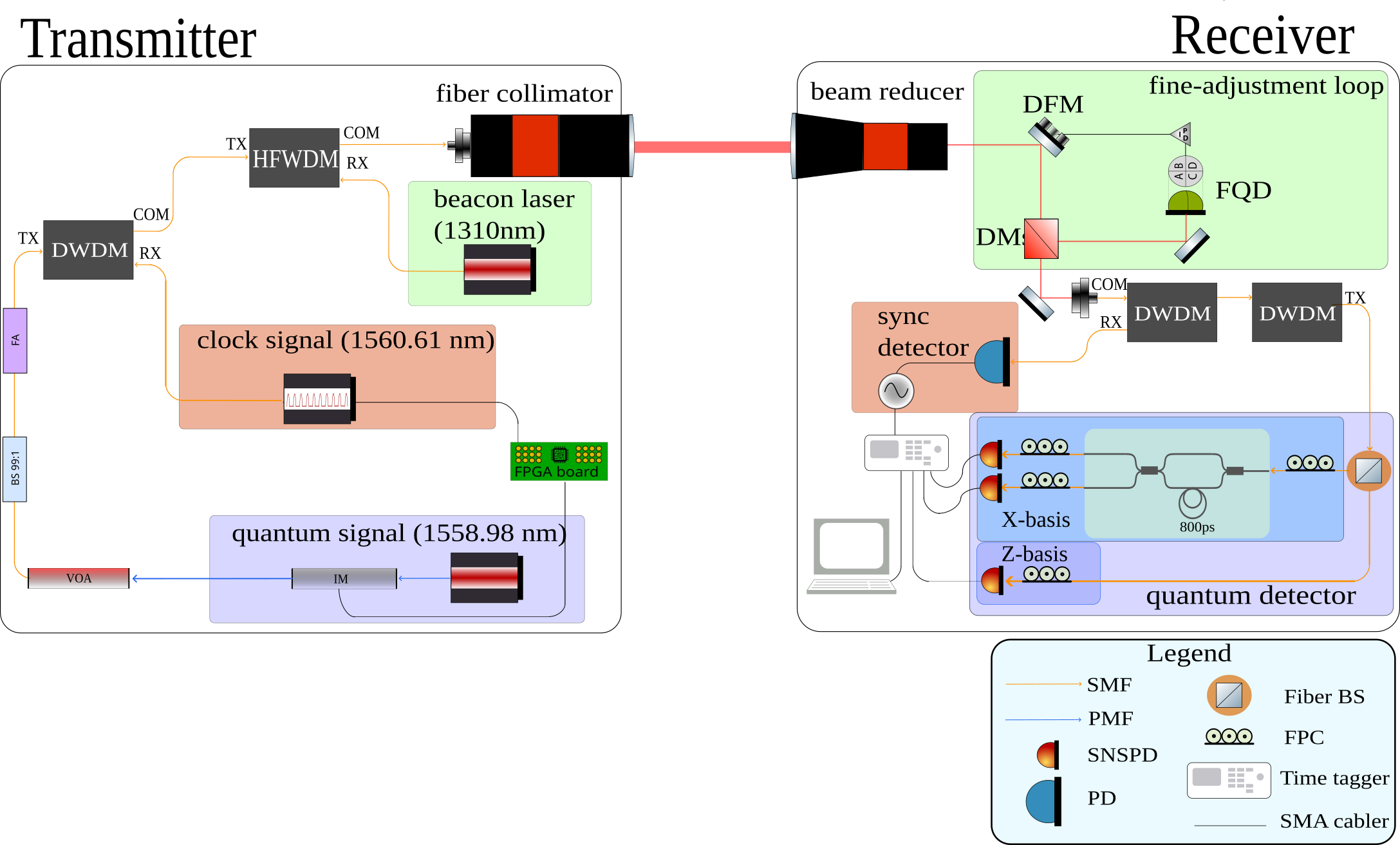}
    \caption{
    \textbf{Detailed setup scheme.}
    FPGA: field programmable gate array;
    IM: intensity modulator;
    VOA: variable optical attenuator;
    FA: fixed attenuation;
    DWDM: dense wavelength division multiplexer;
    HFWDM: high-frequency wavelength division multiplexer;
    SMF: single-mode fiber;
    PMF: polarization maintaining fiber;
    DFM: deformable mirror;
    FQD: four-quadrant detector;
    DM: dichroic mirror;
    CW: continuous wave;
    PD: photodiode;
    FPC: fiber polarization controller.}
    \label{fig:detailed-setup}
\end{figure*}

Figure \ref{fig:detailed-setup} shows the detailed view of the experimental setup used both in the $50\,$m long and in the $500\,$m long link. 
The transmitter generates the QS at $1558.98\,$nm, the CS at $1560.61\,$nm, and the BL at $1310.10\,$nm. 

For the QS, the qubit has a temporal duration of \(1680\,\)ps, and the time difference between the E and L pulses is \(800\,\)ps. 
To produce these pulses, an IM carves the light coming from a $<100\,$kHz narrow band CW laser.
The $V_{\pi}$ can be in range between $-12\,$V and $12\,$V.
Thus, we need a driver circuit to amplify the FPGA signal since it is bound to a maximum value of $1\,$V.
Then, the FPGA provides a \(1.2\)GHz electrical signals sequence which, after being amplified to $V_{\pi}$, allows to carve half states on the X-basis and half on the Z-basis.
Then, to reach the single-photon intensity, a BS 99:1 adds 20 dB attenuation at the 1\(\%\) output port, and the other two fiber attenuators add \(55\,\)dB.

The CS is a train of pulses carrying the FPGA's $10\,$MHz signal clock or a $145\,$kHz signal clock for, respectively, the $50\,$m long and the $500\,$m long link. 
For the $50\,$m link the CS is a Small Form-factor Pluggable (SFP) pulsed laser housed into the FPGA.
In the $500\,$m link, we found the $10\,$MHz too susceptible to the signal-fading. 
Thus, for the $500\,$m link, we adopt a lower synchronization frequency of $145\,$kHz.
After the signals' generation, a DWDM combines the CS with the QS.

The BL is a \(1310\,\)nm continuous wave laser which we use as a reference to adjust for the beam wandering effect at the receiver side.
Then, an HFDWM combines the BL at \(1310\,\)nm with the CS and the QS, which have wavelengths greater than \(1550\,\)nm.
Finally, a variable-focus fiber collimator Thorlabs C80APC-C with a diameter of $42.5\,$mm delivers the three signals through the \(50\,\)m FS channel with an approximate beam diameter of \(15\,\)mm.

The receiver collects the incoming signals with a Galilean telescope Thorlabs GBE10-C with an outer lens diameter of \(35\,\)mm and a magnification factor of $10$. 
Right after, the tip and tilt DFM reflect the beam toward a DM.
The DM separates the BL from the other signals, and an FQD measures the beacon's position in the transverse plane.
The FQD (Thorlabs PDQ30C) has four photodiodes to accurately measure the displacement of an incident beam relative to the calibrated center.
The FQD resolution is \(0.75\,\mu\)m and sensor depth of \(3.05\,\)mm, equivalent of \(0.12\,"\).

The displacement between the beacon position and the center of the FQD determines the feedback error \(\vec{e}_P(t)\).
The vector notation denotes two components in the error, namely the tip and tilt components. 
In the notation, \(P\) stays proportional, and \(t\) for the iteration time.
Then, after estimating at each iteration the integral error \(\vec{e}_I(t)=\sum_{i=0}^{t}\vec{e}_P(i)\) and the derivative error \(\vec{e}_D(t)=\vec{e}_I(t)-\vec{e}_I(t-1)\) we use a PID controller to close the loop on the DFM.
With this fine-adjustment correction, we aim to optimize the coupling of the QS and the CS into the SMF.

The QS and CS arrive in the laboratory following the SMF, traveling through the pre-existing CNR-INO's fiber infrastructure.
In the laboratory, the CS and the QS follow two separate directions after a DWDM.
On the CS path, we retrieve the clock using a PD. 
The clock is essential for the time-tagger unit to label each photon’s click.

Following the QS path, the qubits enter the quantum receiver.
First, a BS with a 50:50 ratio replicates Bob's random basis choice.
An SNSPD measures the arrival times of photons directly from the first output port, which is valid for the Z-basis detection.
The second output port is routed towards an IMZI with a delay line of \(\tau=800\,\)ps in one arm.
The PIC comprises a borosilicate glass matrix, and the waveguides are made through an ion-exchange process \cite{broquin2021integrated}.
SMFs are edge-coupled and glued to the chip.
Due to the birefringence of the matrix, the PIC component needs to be temperature stabilized by a Peltier cell, controlled with a PID loop.
Additionally, since the PIC is polarization-dependent, an FPC is required.

A second SNSPD detects the photon clicks from one output of the IMZI, applicable for X-basis detection.
Then, we calculate the visibility of the interferometer using the formula
\begin{equation}
V = \frac{\max{I}-\min{I}}{\max{I}+\min{I}},
\end{equation}
where $I\propto n$ is the counts rate from the IMZI output, and $\min{I}$ is determined by adjusting the phase of the interferometer.
Since modifying the PIC IMZI phase through temperature adjustments is time-consuming, we utilize another SNSPD after the second IMZI output specifically to measure the minimum interference.
It is worth noting that while this strategy is advantageous in terms of time efficiency, the QKD setup can still be carried out with only two detectors.

Finally, a time-tagger unit converts each SNSPD signal into digital data, which we process using our software.
The software calculates the $QBER_Z$ and the $QBER_X$ from the visibility using the formula
\begin{equation}
    QBER_X = \frac{1 - V}{2}.
\end{equation}
Then, we extract the length of the secure key one can obtain given the measured values of the QBERs and the number of events per second for the selected block size of $n_Z=10^7$.

\end{multicols}

\end{document}